\documentclass[preprintnumbers,amsmath,amssymb,letterpaper]{revtex4}
\usepackage{graphicx}
\pdfoutput=1

\begin{document}

\title{Unveiling the Relationship Between Structure and Dynamics in
Complex Networks}

\author{Cesar H. Comin, Jo\~ao B. Bunoro, Matheus P. Viana and Luciano da F. Costa$^1$}

\affiliation{
	$^1$Institute of Physics at S\~ao Carlos, University of S\~ao Paulo, S\~ao Carlos, S\~ao Paulo,
13560-970 Brazil \\
}

\date{\today}

\begin{abstract}
\end{abstract}

\maketitle

{\bf Over the last years, a great deal of attention has been focused on
complex networked systems, characterized by intricate structure and
dynamics~\cite{barrat:2008,Costa:2011,sporns:2004,barabasi:2004,satorras:2004,castellano:2009}. The
latter has been often represented in terms of overall statistics (e.g. average
and standard deviations) of the time signals~\cite{barrat:2008}.
While such approaches have led to many
insights, they have failed to take into account that signals at
different parts of the system can undergo
distinct evolutions, which cannot be properly represented in
terms of average values. A novel framework for identifying the principal
aspects of the dynamics and how it is influenced by the network structure is
proposed in this work.  The potential of this approach is illustrated with respect
to three important models (Integrate-and-Fire, SIS and Kuramoto),
allowing the identification of highly structured dynamics, in the sense that
different groups of nodes not only presented specific dynamics but also
felt the structure of the network in different ways. }

Given that complex systems are almost invariantly composed by a large
number of interacting elements, they can be effectively represented and
modeled in terms of complex networks~\cite{newman:2003,barabasi:2002}.
With this mapping, their structural and dynamical properties can be extracted
and investigated. Typically, the structure of such networks is quantified
in terms of several measurements~\cite{costa:2007}, reflecting different
properties of the respective topology (e.g. node degree, shortest paths,
centralities) and geometry (e.g. arc length distances, angles, spatial density).

A great deal of the investigations about structure and function in
complex systems has focused on trying to predict the dynamics from
specific structural
features~\cite{barrat:2008,boccaletti:2006,dorogovtsev:2008}.
Such an ability would provides the means for effectively controlling
real-world systems. Despite the growing
number of works devoted to this problem, the knowledge about the
relationship between the structural and dynamical properties remains
incipient because of two main reasons: (a) dynamics is
often summarized in terms of global statistics, overlooking the
intricacies of dynamics; and (b) the investigation often
focuses on linear relationships such as correlations between
structural and dynamical features.

The present work is aimed at addressing these limitations
through a comprehensive and systematic procedure, involving three steps described as following. First, we use
the multivariate statistical approach known as Principal Component
Analysis (PCA)~\cite{Jolliffe:2002} in order to identify the most
important features of the dynamics unfolding at every network node.
Then, we apply a
procedure for checking at what an extent they are determined by \emph{any}
structural property of the system.  Therefore, we first identify the
dynamical features that are affected by the network structure and
thereafter probe the effect of a given set of structural measurements
on those features (see supplementary Figure 1). We found that the
time signals are organized into well-defined patterns (see box for a simple example of structured dynamics),
which is henceforth called \emph{structured dynamics}.

After the time signal $x_i(t)$ at each node $i$ is recorded along $T$
consecutive discrete time instants ($t=1,2,\cdots,T$), the PCA methodology can be applied in order
to obtain linear combinations of the signal components (new random
variables) that can be understood as new measurements $PCA_i^{(1)},
PCA_i^{(2)}, \ldots, PCA_i^{(M)}$ ($M\leq T$) characterizing each
individual signal $i$. These measurements are completely
uncorrelated and correspond to respective projections of the original
space along axes that are optimally aligned along the directions of
maximum variation of the dynamics. Therefore, the PCA provides
a compact and effective description of the original time signals. In case they
are correlated, just a few principal axes are required for explaining the most
relevant aspects of the original dynamics. The present work focuses
attention on this category of dynamical systems.

Having identified the most relevant aspects of the time signals, it is now
important to check at what an extent each of the dynamical variables
$PCA^{(m)}$ is being influenced by \emph{any} of the structural aspects of
the system under study.  Thus, the method relies on no specific
structural measurements whatsoever.  First we assume that the time
signal of a node $i$ can be expressed as in Equation~\ref{eq:sys}.

\begin{equation*}  \label{eq:sys}
	x_i(t+1)=F\left(\left\{x_j(t)\right\}\right),
\end{equation*}

where $j$ indexes the neighbors of $i$, and the initial condition is
$x_i(0) = \xi_i$.  Note that the system
is Markovian, time invariant and that the dynamic function $F$ is assumed to be
identical at each of the network nodes. Therefore, the dynamics at a node
$i$ depends on three elements: (i) the function $F$, (ii) the initial condition
$\vec{\xi}$, and (iii) the topology of the network, given by the
adjacency matrix $\mathbf{A}$.  The null hypothesis is that the time
signals cannot be discriminated one another.  The given network is
subjected to a large number of simulations starting with different
initial conditions $\vec{\xi}$. Then, the density functions $P(PCA^{(m)})$
for each principal dynamical feature $PCA^{(m)}$ are estimated and used
as references.  Thus, each of these density functions provides an
estimate of the values obtained for the feature $PCA^{(m)}$
considering all the network nodes.  Also, the individual density
functions $P_i(PCA_i^{(m)})$ of the values of $PCA_i^{(m)}$ are estimated
considering each node $i$ individually. In order to test whether the structural features
produce distinct dynamical behavior at node level, we introduce the
index, $\alpha_i^{(m)}$, between $P(PCA^{(m)})$ and $P_i(PCA_i^{(m)})$,
which is defined as the Euclidean distance between the respective
densities.  In other words, the value of $\alpha_i$ expresses how
much the node $i$ deviates from the null hypothesis.  In
case $\alpha_i^{(m)}$ is significantly small, the dynamical property
$PCA_i^{(m)}$ at node $i$ is understood not to be discriminated by any
structural features, confirming the null hypothesis.

For each dynamical feature $PCA^{(m)}$ for which the null hypothesis is
not verified, we then quantify
how much it is related to each of a predefined set of structural
measurements $s^{(1)}, s^{(2)}, \ldots, s^{(S)}$. To do so, we calculate the
dispersion of $\alpha$ along the measurement, weighted by its density. This is
conveniently given by the conditional entropy (or
equivocation~\cite{Arndt2004}) between $s$ and $\alpha$.

In order to illustrate the proposed approach, we resort to three well-known
types of dynamics in complex systems, namely
integrate-and-fire~\cite{guardiola:2000,roxin:2004}, epidemics
spreading (SIS)~\cite{pastor-satorras:2001a,pastor-satorras:2001b},
and synchronization
(Kuramoto oscillators)~\cite{hong:2002,moreno:2004,arenas:2008} (see the Box).
Although the proposed methodology can be applied to any regime,
we consider the steady-state values of the signals in the Erd\H{o}s-R\'{e}nyi (ER)
model.

We start by investigating the \emph{integrate-and-fire} dynamics in which each
node is understood as a neuron undergoing the McCulloch and Pitts
integrate-and-fire dynamics~\cite{koch:1999,burkitt:2006}.  The binary
time signal (firing spikes) of each node is recorded and represented as a vector.
PCA is then applied over this ensemble of vectors to project the data onto the
eigenvectors associated with the three largest eigenvalues.
Figure~\ref{f:projs}(a) shows the PCA projections of the time signals,
with the colors corresponding to the node degrees (see also the supplementary
Figure 2). Remarkably, the nodes
resulted organized according to a well-defined geometrical pattern.  Also shown in
Figure~\ref{f:projs} are the densities (Figs.~\ref{f:projs}b,d)
and average entropies (Figs.~\ref{f:projs}c,e) of the original
signals after being further projected into two dimensions.
 Note that the 2D PCA projections include groups of nodes organized
 as `cords' (see supplementary Figure 3).
 The obtained 3D projection
can be divided into three main regions: (i) a relatively sparse conic `head';
(ii) a densely populated `waist'; and (iii) a relatively dense and complex `tail'.
We observe that the density peaks occur at low entropy places.  This
suggests that the system dynamics tends to unfold so as to favor more
ordered (i.e.\ lower entropy) time signals.

Another dynamical process considered in this paper is the
\emph{Susceptible-Infected-Susceptible} (SIS) model
\cite{Satorras2001} that is used to investigate epidemics spreading on
networked systems. For this dynamical process, the time series
associated to each node is also binary. Figure~\ref{f:projs}(f) shows the PCA
projection of the SIS dynamics (node degrees mapped into colors), as
well as the respective further two-dimensional projections into the
$PCAö{(1)} \times PCAö{(2)}$ plane. The resulting PCA pattern is much
simpler than that obtained for the integrate-and-fire dynamics, now containing
just an eye-shaped volume. A cord can also be identified in this projection
(see supplementary Figure 3). A similar trend is
observed  between the time signal density and entropy (Figure~\ref{f:projs}g,h).

The third dynamics investigated in this work is the \emph{Kuramoto
  dynamics}~\cite{moreno:2004}, whose PCA projection is shown in
Figure~\ref{f:projs}(i), with the respective one-dimensional
projection given in (j). It is clear from these results that the
Kuramoto dynamics yielded the simplest PCA
projection, in the sense that most of the signals variance is
explained by $PCAö{(1)}$.  We also found that the Kuramoto
dynamics tends to yield cords which are related to the natural
frequencies of the nodes, parametrized by the respective relative
phases (see supplementary Figure 4).

In the second step of our methodology, we evaluated the
$\alpha_i^{(m)}$ for each of the new measurements $PCA^{(m)}$,
as illustrated in Figures \ref{fig:alphas}.  The statistical significance of
the $\alpha$ values was obtained by using a random null model
(see supplementary Figure 5). First, several simulations
were performed for varying random initial conditions $\vec{\xi}$, from
which the reference histograms $P(PCA^{(1)})$, $P(PCA^{(2)})$ and $P
(PCA^{(3)})$) in Figure \ref{fig:alphas}(a-c) were obtained for the
integrate-and-fire dynamics. These densities were also estimated for each node
separately. For instance, in Figure \ref{fig:alphas}(d-f), (g-i), and (j-l), we illustrate
the densities of $PCA^{(1)}$, $PCA^{(2)}$ and $PCA^{(3)}$, respectively,
for three randomly chosen nodes. The small
values of $\alpha$ obtained in (g) indicate that this nodes
is not being indistinctly influenced by any structural features of the network. On the other
hand, larger values of $\alpha$, such as those in (f,i,k,l) corroborate
that the time signals at those respective nodes are greatly affected by
the network structure. The overall $\alpha$ densities are
shown in Figure \ref{fig:alphas}(m-o) for integrate-and-fire, (p,q)
for SIS, and (r) for Kuramoto.  Remarkably, the wide dispersion of
$\alpha$ values obtained for most cases confirms that the structural
influence on the dynamics can vary strongly from one node to another.
It is also clear from these results that the $PCA^{(3)}$ of the
integrate-and-fire dynamics, as well as the $PCA^{(2)}$ of the
SIS dynamics are the dynamical features mostly affected by the
topology of the network.  As expected, because of the adopted strong
coupling, the Kuramoto dynamics resulted to be largely independent of
the network structure, which was duly identified by the proposed
methodology. This is confirmed by the rather small values of $\alpha$
shown in the density in Figure \ref{fig:alphas}(r).  A stronger influence of
the topology was verified for a weaker coupling (see supplementary
Figure 6).

We now proceed to the third step, obtaining scatterplots of the $\alpha^{(m)}$
versus some topological measurements.  The respective conditional entropies
are then estimated in order to quantify the degree at which the $\alpha$ values are
being explained by each of the measurements. The scatterplots respectively to
the smallest conditional entropies (see Table~\ref{t:centropy}) are depicted in Figure~\ref{f:centropy}.  In
case of the integrate-and-fire dynamics (Fig.~\ref{f:centropy}a-c),
we have that the eigenvector centrality was the topological feature
that most strongly affected all the three principal component
variables. As can be be seen in the scatterplot for $\alpha^{(3)}$ versus
$EC$ (Fig.~\ref{f:centropy}c), the influence of the topology on the
$PCA^{(3)}$ measurement is felt more strongly for the nodes with lower $EC$ values.
As shown in Figure~\ref{f:centropy}(d,e), respective to the SIS
dynamics, the degree was the measurement most directly related to the
two principal variables.  Other measurements also were found to be
related to the 2D PCA projections (see supplementary Figure 7).
Remarkably, the values of $\alpha^{(1)}$ and
$\alpha^{(2)}$ present a low peak at values of degree similar to the average
network degree ($\langle k \rangle=10$), meaning that both $PCA^{(1)}$ and
$PCA^{(2)}$ variables of the SIS tend to feel the topology
only for nodes with degree distinct from $\langle k \rangle$.  Although the
Kuramoto dynamics resulted unaffected by the network structure, the
accessibility was the topological feature that yielded the smallest conditional
entropy.

All in all, we have proposed a new methodology for investigating the
relationship between structure and dynamics in complex networked
systems. The reported approach relies on two critical concepts,
namely the consideration of the structure/dynamics relationship at the
individual node level and also along different values of specific
structural measurements. These concepts allowed the separation of the
intermixing effects that would be otherwise obtained by using
traditional approaches where both structure and dynamics are
summarized in terms of global statistics. The obtained results
corroborated the validity and importance of these hypotheses. Moreover,  despite
the uniformity of the ER topology, we identified
a highly structured dynamics. In the case of the integrate-and-fire and the SIS
dynamics, we found that the PCA regions with higher density of nodes
tended to present low signal entropy, which suggests that the dynamics
is related to signal uniformity.

\section{Supplementary Material}

\subsection{Diagram}

\begin{figure}[!h]
\begin{center}
  \includegraphics[width=0.5\linewidth]{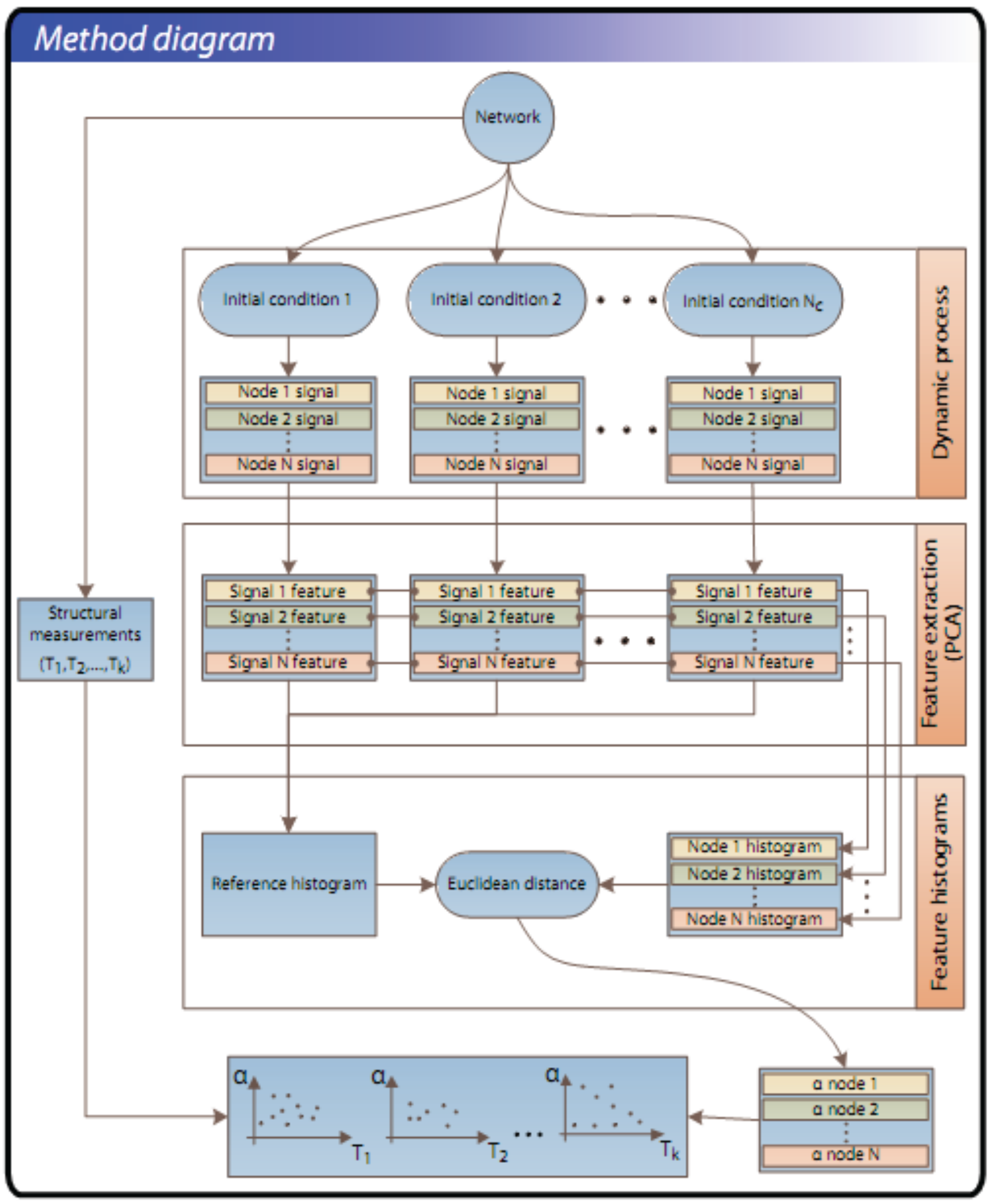}
  \caption{Flow diagram of the framework proposed in this article
  for investigating the relationship between structure and dynamics
  of complex systems.}
 \label{fs:diagram}
\end{center}
\end{figure}

The framework proposed in the present article for investigating
the relationship between structure and dynamics in complex
systems is summarized by the flow diagram in Figure~\ref{fs:diagram}.
First, during the simulation stage, the original network is
subjected to a total of $N_C$ random initial conditions and
the time signals for each node is recorded after the steady
state has been reached.  The next step involves the extraction of
PCA dynamics features from the time signals.  These features are
then used to obtain the respective reference histogram.
The Euclidean distances between this reference
histogram and the node histograms are obtained and used to
estimate the $\alpha$ values for each node with respect to each
dynamical feature.  Measurements $ T_i$ of the structure of the
network are also estimated and used to obtain the scatterplots
$\alpha \times T_i$.

It should be observed that the orientation of the PCA axes
is undetermined, because the negative of an eigenvector is
also an eigenvector associated to the same eigenvalue.
Therefore, in order to provide a stable reference, we obtained
a set of eigenvectors for the large network (10000 nodes) and
adopted then for all subsequent projections.

\newpage

\subsection{Integrate-and-Fire time signals}

\begin{figure}[!h]
\begin{center}
  \includegraphics[width=0.8\linewidth]{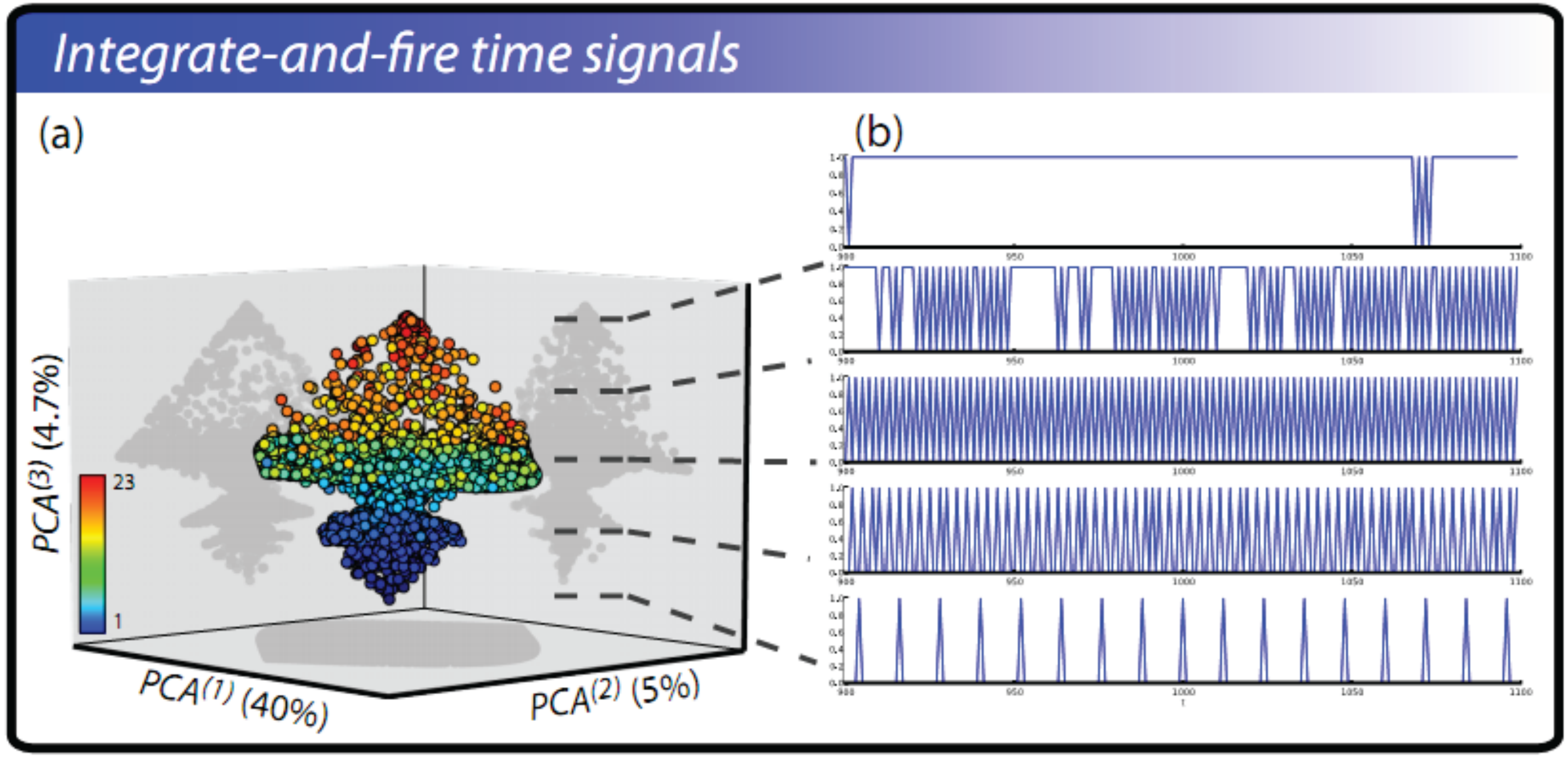}
  \caption{(a) The 3d PCA projections of the integrate-and-fire dynamics
                unfolding in an ER network. (b) Examples of time signals
                obtained for the stratifications of the dynamics along the
                $PCA^{(3)}$ axis.  Observe that the frequency of the signals
                tends to decrease with lower values of $PCA^{(3)}$.}
 \label{fs:signals}
\end{center}
\end{figure}

The integrate-and-fire dynamics has been extensively investigated in
complex systems because not only of its biological inspiration (i.e.\ as
a simplified model of neuronal networks), but also as a consequence of its
rich behavior~\cite{barrat:2008,Costa:2011,sporns:2004,barabasi:2004,satorras:2004,castellano:2009}.  Though initial investigations focused on more
regular topologies such as layered systems and lattices~\cite{newman:2003,barabasi:2002}, much
attention has been driven to the study of integrate-and-fire unfolding in complex
networks~\cite{koch:1999,burkitt:2006}.  Given that different network models are
characterized by specific topological features, a fundamental question
arises regarding the relationship between such structural properties and
the respective dynamics. Remarkable related results have been obtained
through the application of
the methodology proposed in this article.  Figure~\ref{fs:signals} shows the
three-dimensional PCA representation of the integrate-and-fire dynamics
unfolding in an ER network with 10000 nodes and average degree 10.
The integrate-and-fire realization considered initial conditions drawn uniformly
within the range $[0,\tau+1]$, with $\tau=8$.
The time signals were recorded along 1000 steps after the system had
reached steady state.

The nodes were colored according to their degree, which resulted stratified
along the $PCA^{(3)}$ axis.  Examples of respective time signals are shown
on the righthand side of Figure~\ref{fs:signals}.  Interestingly, the time signals
tended to be distributed along the $PCA^{(3)}$ axis in terms of their respective
average frequency. The top of the PCA structure is populated by signals with maximum
frequency, i.e.\ a firing spike at each time instant.  The frequency decreases
as one moves downwards along the $PCA^{(3)}$ axis, such that the lowest
frequency signals are found at the bottom (tail) of the PCA structure.  The largest
number of time signals (about 40\%)  were found to have average period of 2 cycles
and to be distributed along the `waist' of the PCA structure.  As it becomes clear
from Figure 3c, the time signals tended to be organized in groups with
similar frequencies corresponding to integer number of cycles, in a way that reminds
harmonic frequency distribution in a linear system.  This finding implies that
systems undergoing integrate-and-fire dynamics with similar parameters will
have most neurons firing at 2 cycles, with the others tending to have multiple
integer periods of oscillations.  This interesting finding is related to the fact that
the investigated dynamics in the ER structure tends to favor the uniformity
of the time signals, as reflected in the respective entropy.  Particularly
remarkable is that such structured dynamics, including marked clusters of time
signals, arose despite the structural uniformity of the ER network.

\newpage

\subsection{Cords}

\begin{figure}[!h]
	\includegraphics[width = 0.8\linewidth]{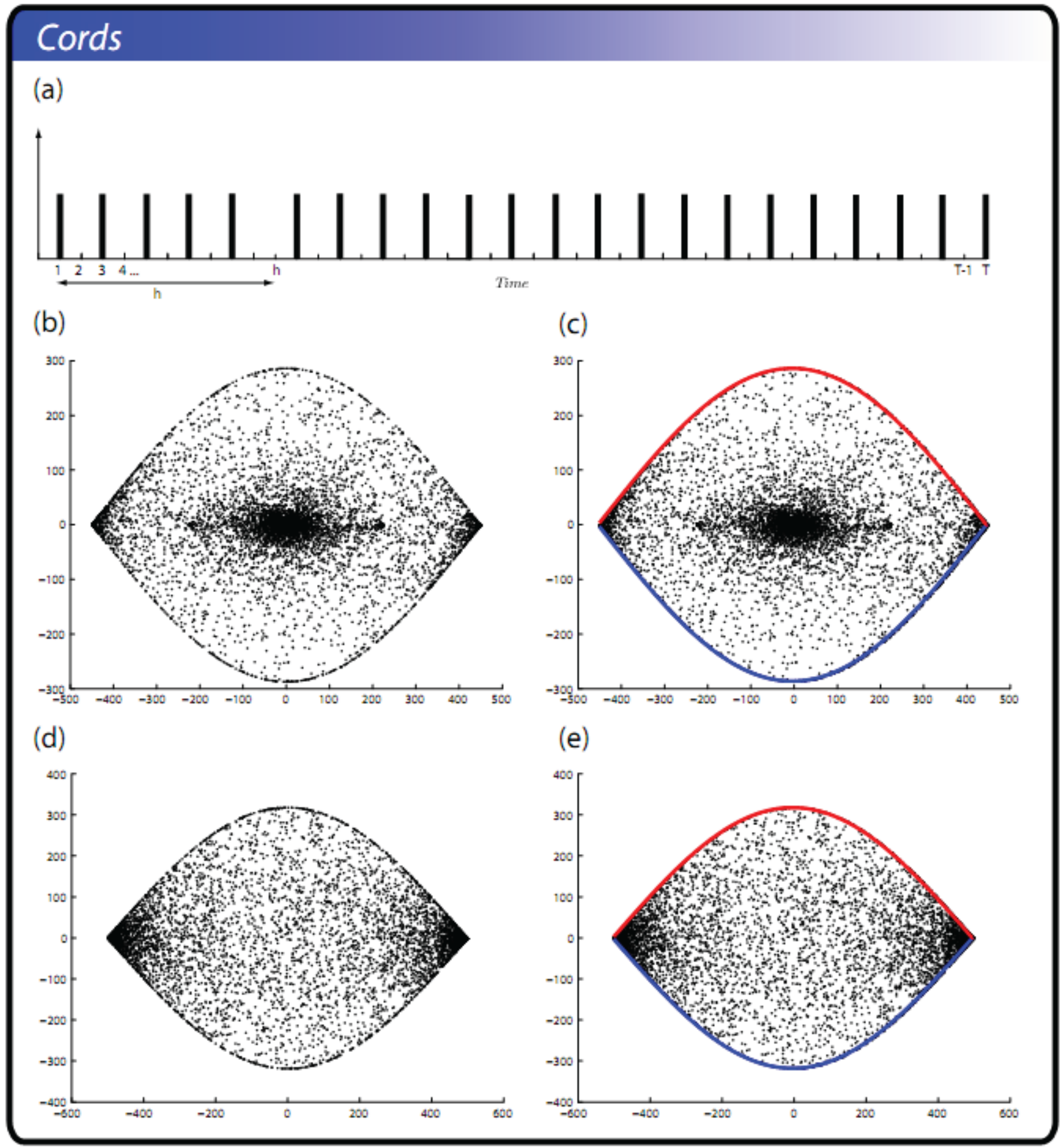}
	\caption{(a) General form of the one-hole (two consecutive zeros) time signal.
	             The 2D PCA projection of the (b) integrate-and-fire dynamics and (d) SIS.
	              The same, (c) and (e) respectively, with the analytical parametric
	              curves approximating the cords}
	\label{f:cords}
\end{figure}

In order to explain the organization of signals along the cords in the PCA projections,
we develop now a simple formulation for the values of $PCA^{(1)}$ and $PCA^{(2)}$,
which we will call $x$ and $y$ for simplicity. First, let us assume we have the signal
showed in Figure \ref{f:cords}(a), which can expressed as

\begin{equation}
s(t) = \sum_{i=1}^N \delta_{t,v_i},
\label{eq:signal}
\end{equation}

where, $N$ is the total number of spikes in the considered time interval, the $i-$th element
of the $\vec{v}$ identifies the instant of the $i-$th spike.
We can write the following rule for the elements of $\vec{v}$

\begin{equation}
v_i =
\left \{
\begin{array}{rcll}
2i-1 & \text{if } 1 \leq i < h\\\\
2i & \text{if } h \leq i \leq N.
\end{array}
\right.
\end{equation}

On the other hand, the first and second eigenvectors of the \emph{integrate-and-fire} and
SIS dynamics can be expressed as

\begin{equation}
e_1(t) = (-1)^t
\end{equation}

and

\begin{equation}
e_2(t) = (-1)^t\cos\left(\frac{t\pi}{T}\right),
\end{equation}

where $1\leq t \leq T$, and $T$ is the size of the time window. Therefore, we can
estimate the values of $x$ and $y$ considering the inner-product of the signal and the respective eigenvector:

\begin{equation}
x = \sum_{t=1}^T(-1)^ts(t)
\end{equation}

and

\begin{equation}
y = \sum_{t=1}^T(-1)^t\cos\left(\frac{t\pi}{T}\right)s(t).
\end{equation}

Substituting Equation \ref{eq:signal}, we find the PCA projections as a function of the hole position, $h$

\begin{equation}
x(h) = N - 2(h+1)
\end{equation}

and

\begin{equation}
y(h) = \sum_{i=h}^N \cos\left(\frac{2i\pi}{T}\right) - \sum_{i=1}^{h-1} \cos\left(\frac{(2i-1)\pi}{T}\right).
\end{equation}

In Figure \ref{f:cords}(b) and \ref{f:cords}(c) we show the plot of $y$ in terms of $x$ (blue curves)
for different values of $h$ considering the integrate-and-fire and SIS dynamics, respectively. It is
interesting to observe that if we consider the first element of the time signal as being zero
(instead of one), the resulting curves corresponds to the red curves.

\newpage

\subsection{Phases in Kuramoto}

\begin{figure}[!h]
	\includegraphics[width = 0.8\linewidth]{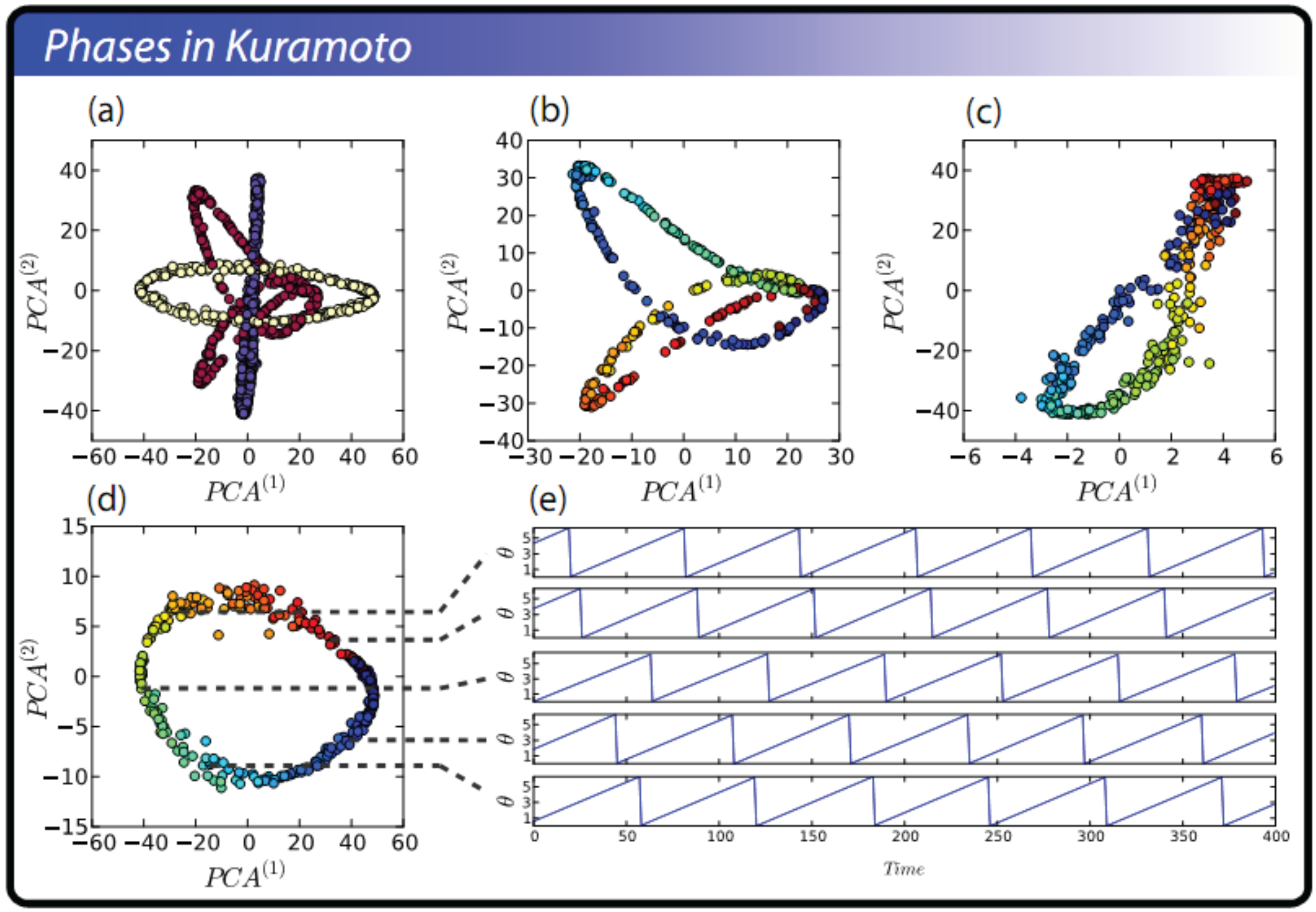}
	\caption{Illustration of the distribution of phases of the time signals along the cords
	              yielded by the Kuramoto dynamics for an ER network with three
	              frequency groups.}
	\label{f:phase}
\end{figure}

A different parametric configuration of the Kuramoto dynamics was used in order to
further illustrate the potential of the PCA methodology in identifying meaningful
properties of the dynamics.  Three groups of nodes in an ER network with 1000
nodes were assigned specific natural frequencies, with weak coupling ($\lambda=0.5$).
Figure~\ref{f:phase}(a) depicts the 2D PCA projections obtained for the respective
dynamics.  The colors in this figure identify each of the frequency groups.  Their
respective projections are also shown in Figures~\ref{f:phase}
(b-d).  The colors in these three figures identify the relative phase of the
time signals at each node.  Figure~\ref{f:phase}(e) illustrates five characteristic time
signals obtained along the cord in Figure~\ref{f:phase}(d).

\newpage

\subsection{Statistical Significance of $\alpha$}

\begin{figure}[!h]
\begin{center}
  \includegraphics[width=0.6\linewidth]{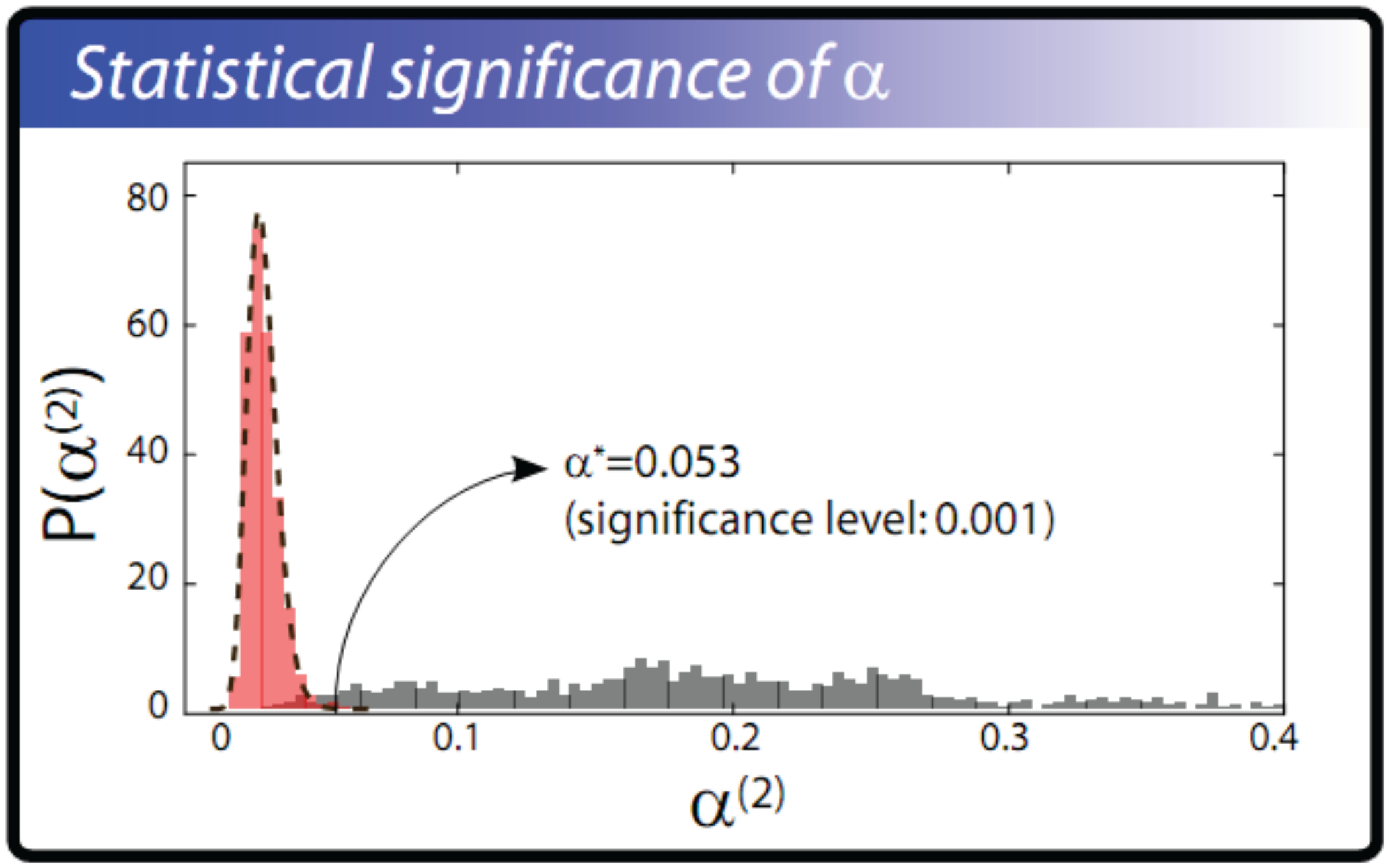}
  \caption{Illustration of the construction used to obtain levels of
        statistical significance for the values of $\alpha$.}
  \label{f:sigalpha}
\end{center}
\end{figure}

In order to obtain a level of statistical significance for the values of $\alpha$
produced by our simulations, we also consider the distribution of $\alpha$ values
that would be obtained in the case of a random null model.  In this model
the value of $\alpha$ for each node was obtained by sampling, through Monte
Carlo, $N_C$ PCA features from the reference histogram. Figure~\ref{f:sigalpha}
illustrates this approach with respect to the $PCA^{(2)}$ variable in
 integrate-and-fire dynamics.  The histogram for the null model, shown in red,
is well-fitted by a log-normal distribution.  The histogram of $\alpha$ values
obtained for the integrate-and-fire configuration is shown in gray.  By
comparing the latter histogram with the fitted log-normal
distribution, it is possible to calculate the significance level assuming 0.001
confidence.  In this specific case, we obtained $\alpha^* = 0.053$, which
corresponds to the probability of obtaining a value of $\alpha$ larger than
$\alpha^*$ by chance.

\newpage

\subsection{Effect of coupling in the Kuramoto dynamics}

\begin{figure}[!h]
\begin{center}
  \includegraphics[width=0.6\linewidth]{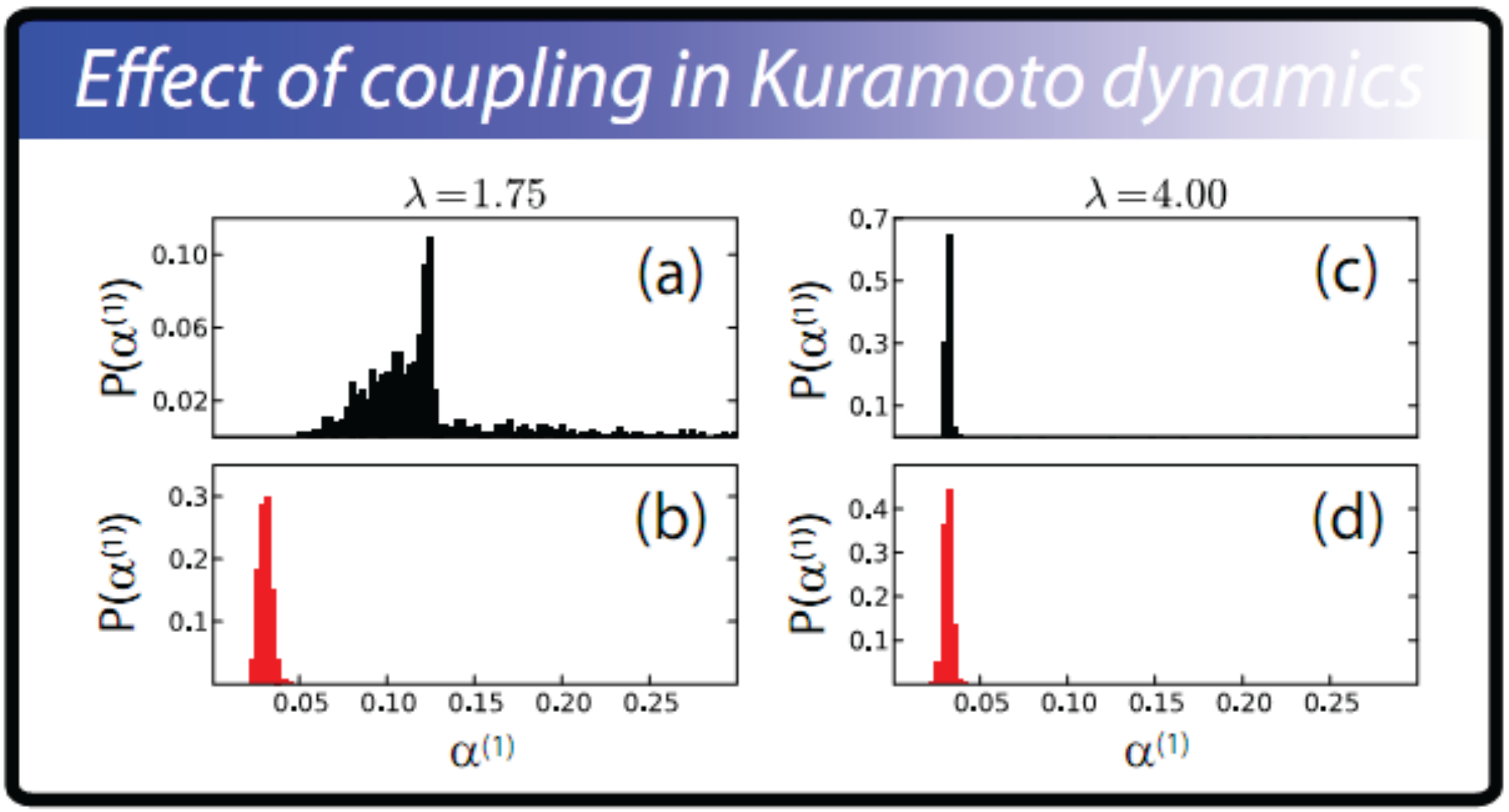}
  \caption{Effect of the coupling in the Kuramoto dynamics.
      As expected, a strong coupling makes the dynamics at each
      node not to be differentiated by any structural features of the network.}
  \label{fs:kuramoto}
\end{center}
\end{figure}

To illustrate the proposed methodology, we considered in the main
article a strongly coupled version of the Kuramoto oscillator.  Here we provide
complementary results with respect to a less intensely coupled configuration.
Figures~\ref{fs:kuramoto}
(a) and (c) show the histograms of $\alpha$ obtained for this type of dynamics
considering relatively weak ($\lambda = 1.75$) and strong ($\lambda = 4.00$)
couplings.  The histograms of $\alpha$ obtained for the null reference model
are shown in Figures~\ref{fs:kuramoto} (b) and (d), respectively.  It is clear from these
results that the values of $\alpha$ are substantially higher than those obtained for the
null model in the case of the less strongly coupled Kuramoto
simulations, while being undistinguishable in the more strongly coupled case.
This means that in the less strongly coupled Kuramoto configuration the dynamics at
each node is differentiated by the network structure, which is not observed in
the other configuration.

\newpage

\subsection{Topological measurements over the 2D PCA projection of the SIS dynamics}

\begin{figure}[!h]
\begin{center}
  \includegraphics[width=0.7\linewidth]{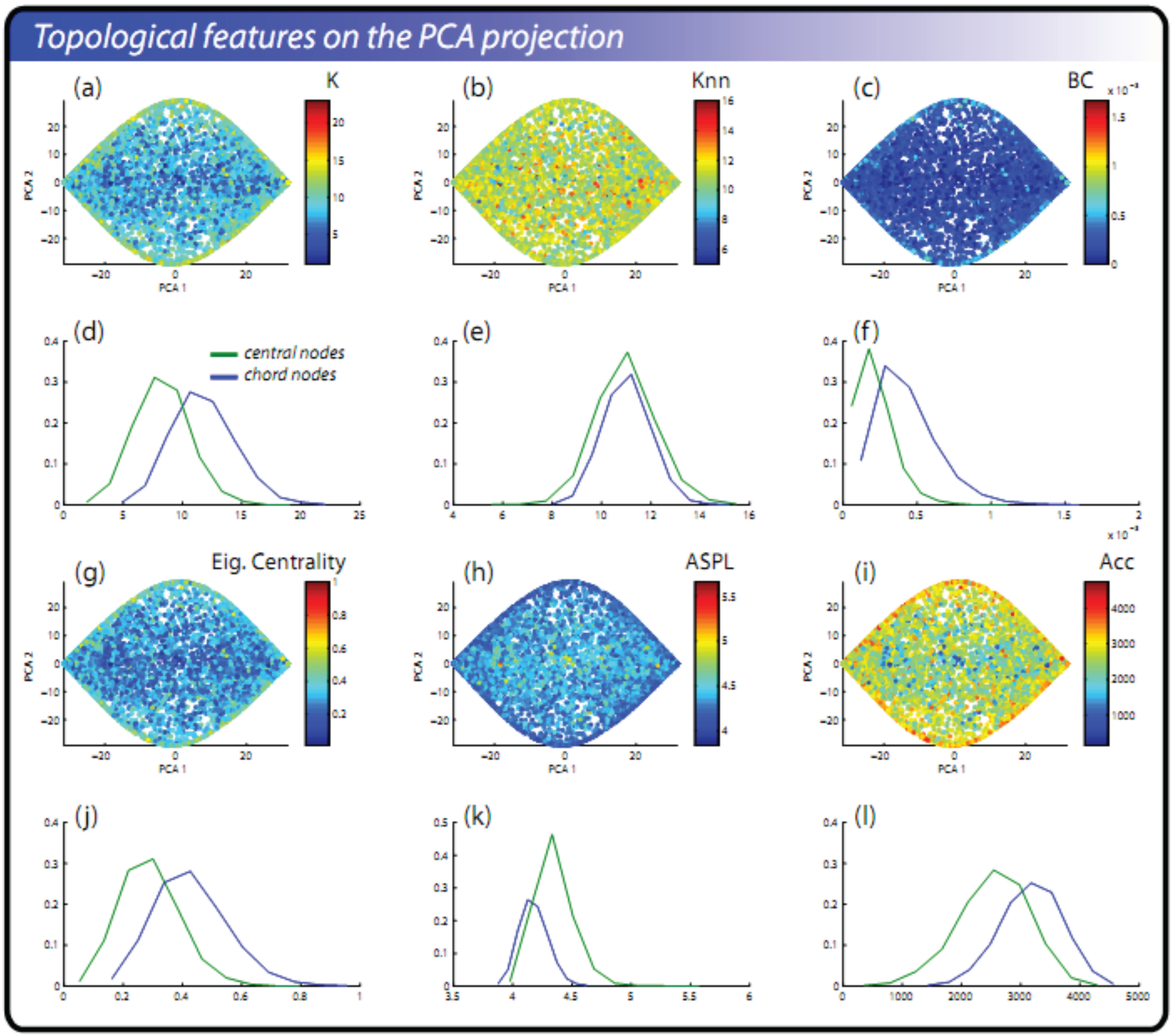}
  \caption{2D PCA projections of the time signals obtained for the
      SIS model colored in terms of the values of specific topological
      measurements (a-c, g-i).  Histograms of the values of the topological
      measurements considering the time signals separation into
      chord and interior regions (d-f,j-l).}
      \label{fs:sis}
\end{center}
\end{figure}

So as to better understand the structure versus dynamics relationship in the
SIS model, we mapped the values of the six considered specific topological
features onto the 2D PCA projections obtained for this
model, as shown in Figure~\ref{fs:sis}.  In each case, the projected
points were separated into two main subsets: one corresponding to the border of
the eye-shaped pattern (a `chord') and the other to the remainder region.
The histograms of the measurements for these two groups are also shown
in Figure\ref{fs:sis} respectively to each 2D projection.  It is clear from these
results that the time signals in these two groups tend to be well-separated
with respect to their degree, eigenvalue centrality, and accessibility.
More specifically, nodes with high values of these three measurements
tend to be found along the chord.  Given that low accessibility values have been
found to be associated to the border of complex networks~\cite{viana2009}, it is possible
that the interior of the eye-shaped regions are occupied by the border nodes
of the ER network adopted in our simulations.

\section*{Acknowledgments}

Luciano da F. Costa is grateful to FAPESP (05/00587- 5) and CNPq
(301303/06-1 and 573583/2008-0) for the financial support. M. P. Viana
was supported by a FAPESP grant (proc. 07/50882-9); J. L. B. Batista
thanks CNPq (131309/2009-9) for sponsorship; and C. H. Comin is
grateful to CAPES for his grant. The authors thank L. Baccal\'a
for his remarks on clustered dynamics, and to G. Travieso and
O. N. Oliveira for commenting on this work. The authors also thank
L. Antiqueira for help with box and for reading and commenting on the manuscript.

\newpage

\begin{table}[h]
\begin{center}
\begin{tabular}{|l|c|c|c|c|c|c|}
\hline
& \multicolumn{3}{|c|}{Integrate-and-Fire} & \multicolumn{2}{|c|}{Susceptible-Infected-Susceptible} &
\multicolumn{1}{|c|}{Kuramoto}\\
\hline
\hline
Measurement & $PCA^{(1)}$ & $PCA^{(2)}$ & $PCA^{(3)}$ & $PCA^{(1)}$ & $PCA^{(2)}$ & $PCA^{(1)}
$ \\
\hline
\hline
$k$ 				& 2.52 & 2.67 & 2.24 & 1.28 & 1.51 & 0.15\\
$knn$ 				& 2.70 & 2.81 & 2.47 & 2.69 & 2.95 & 0.15\\
BC 		& 2.50 & 2.62 & 2.22 & 1.62 & 1.95 & 0.15\\
EC 		& 2.40 & 2.56 & 2.11 & 2.01 & 2.38 & 0.14\\
ASPL 				& 2.43 & 2.60 & 2.14 & 2.00 & 2.37 & 0.13\\
ACC 		& 2.43 & 2.60 & 2.15 & 2.12 & 2.48 & 0.13\\
\hline
\end{tabular}
\end{center}
\caption{The conditional entropies obtained for $\alpha$ considering the
  three dynamics in terms of some topological measurements: degree $k$,
  average neighbor degree $knn$, betweenness centrality $BC$, eigenvector
  centrality $EC$, average shortest path length $ASPL$ and accessibility $ACC$.}
\label{t:centropy}
\end{table}

\newpage

\begin{figure}
\begin{center}
  \includegraphics[width=0.8\linewidth]{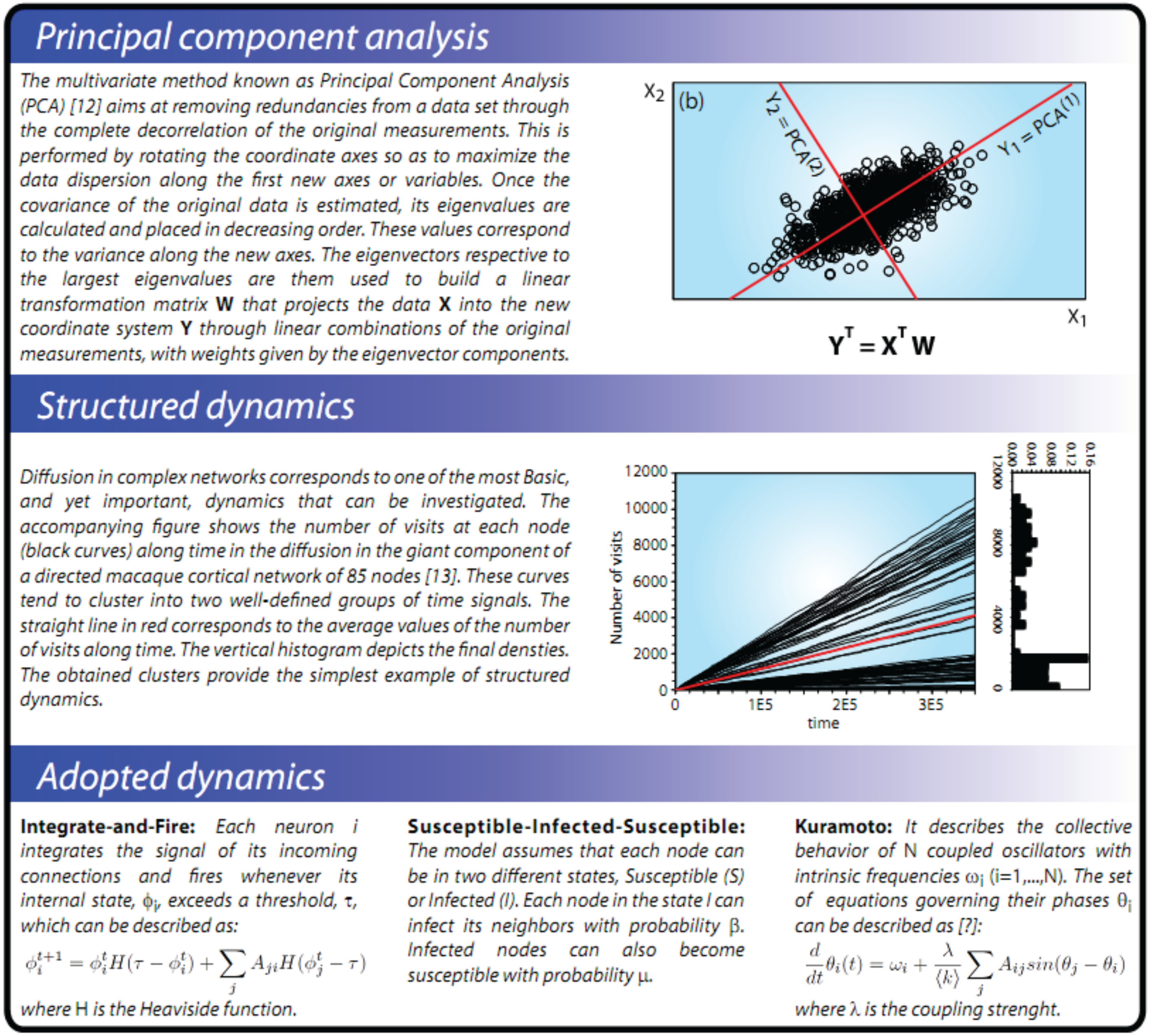}
\end{center}
\end{figure}

\begin{figure}
\begin{center}
  \includegraphics[width=0.8\linewidth]{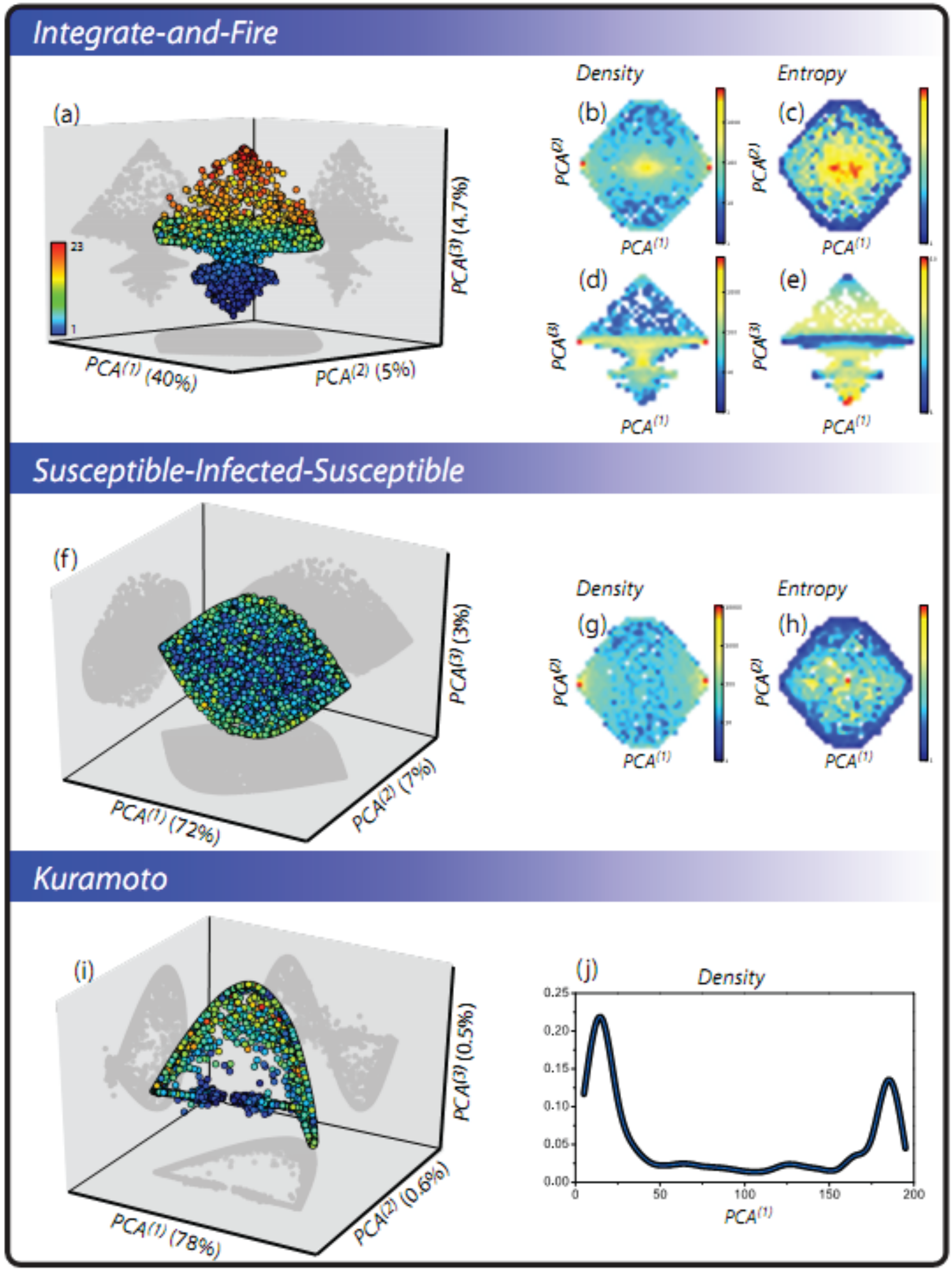}
  \caption{\emph{The 3d PCA projections of the considered dynamics}. (a)integrate-and-fire,
	(f)SIS, and (i)Kuramoto.  The percentage of explained
    variance provided by each principal component variable are shown
    along each axes.  The densities of the projections are shown in
    (b,d) for the integrate-and-fire, (g) for the SIS, and (j) for the
    Kuramoto.  The respective signal entropies are depicted
    in (c,e) for the integrate-and-fire, (h) for the SIS. All these
    results were obtained for an ER network with 10000 nodes and
    average degree of 10.  The integrate-and-fire realization
    considered initial conditions drawn uniformly within
    the range $[0,\tau+1]$, with $\tau = 8$.  The SIS dynamics assumed
    $\beta = 0.8$ and $\mu =1$. The initial condition was such that
    $50\%$ of the nodes were infected.  The Kuramoto simulations were
    performed for $\kappa = 4$, with the natural frequencies drawn
    from a normal distribution with null mean and unit variance and
    the initial phases distributed uniformly between $[0,2\pi]$.
    As shown in (b) and (d), the two extremities of the
    eye-shaped waist corresponds to approximately 40\% of the nodes in the network.
    A third peak is found at the center of the `eye', containing about 22\% of the
    nodes. There is a fourth density peak, located along the central axis of the
    `tail' part of the projection. Likewise, in (g) the time signals concentrate at the two
     extremities of the `eye', corresponding to about 55\% of the nodes.  The number
     of PCA axes considered were either enough to account for at least 75\% of the
     variance or limited to 3. }
\label{f:projs}
\end{center}
\end{figure}

\begin{figure}
\begin{center}
  \includegraphics[width=0.8\linewidth]{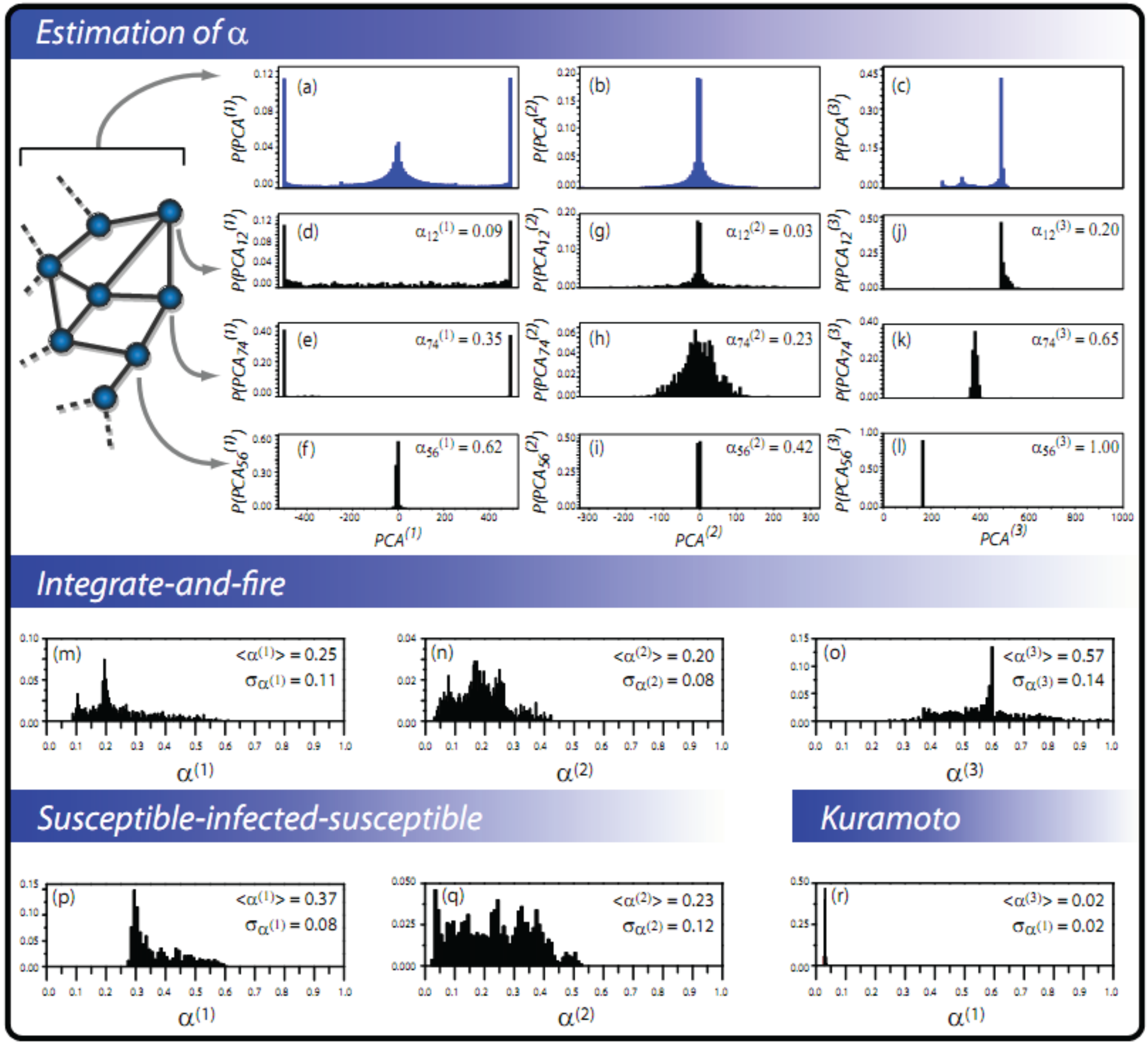}
  \caption{\emph{Estimation of the distribution of $\alpha$}. The reference histograms of (a) $PCA^{(1)}$, (b) $PCA^{(2)}$ and (c)
    $PCA^{(3)}$ for 1000 realizations of the integrate-and-fire dynamics
    over the ER network with 1000 nodes and average degree 10.
    Examples of histograms (considering the same realizations) for
    specific nodes with respect to the (d-f) $PCA^{(1)}$, (g-i) $PCA^{(2)}$,
    and (j-i) $PCA^{(3)}$. The histograms of (m) $\alpha^{(1)}$, (n) $\alpha^{(2)}$
  and (o) $\alpha^{(3)}$ for the integrate-and-fire dynamics; (p)
  $\alpha^{(1)}$, (q) $\alpha^{(2)}$ for the SIS dynamics; and (r) $\alpha^{(1)}$
  for the Kuramoto dynamics.  We used the same set of parameters as
  in Figure~\ref{f:projs}.}
 \label{fig:alphas}
\end{center}
\end{figure}

\begin{figure}
\begin{center}
  \includegraphics[width=0.8\linewidth]{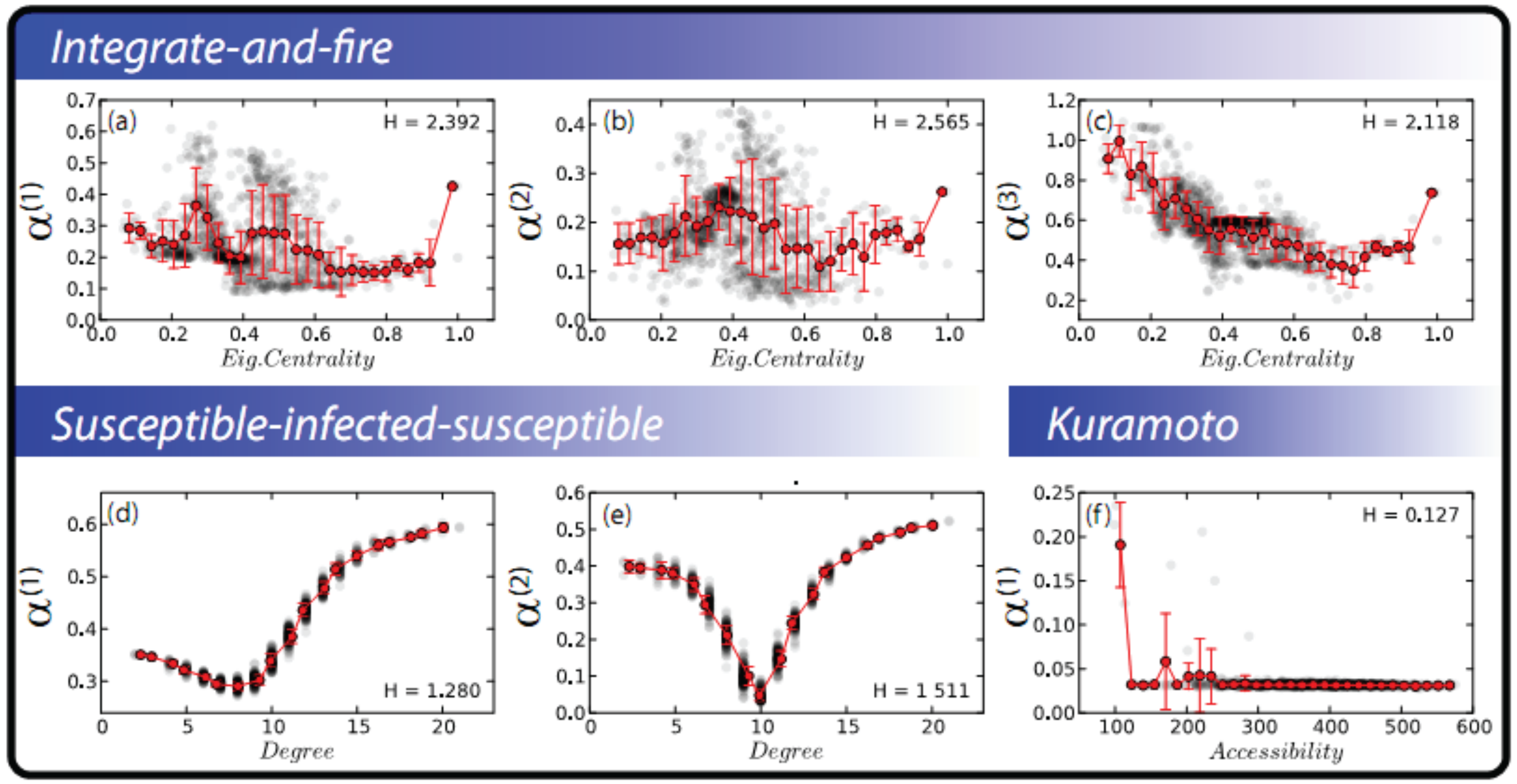}
  \caption{\emph{Scatterplots between topological measurements and
      values of $\alpha$:}  (a) $\alpha^{(1)} \times EC$, (b) $\alpha^{(2)}
    \times EC$,  and (c) $\alpha^{(3)} \times EC$ for the
    integrate-and-fire dynamics;  (d) $\alpha^{(1)} \times k$, and (e)
    $\alpha^{(2)} \times k$ for the SIS dynamics; and (f)
    $\alpha^{(1)} \times Acc$ for the Kuramoto dynamics. The red curve
    corresponds to the average of the $\alpha$ values, and bars
    indicate the standard deviation. Only the scatterplots obtained
    for the smallest conditional entropies in Table~\ref{t:centropy}
    are shown.}
 \label{f:centropy}
\end{center}
\end{figure}

\newpage

\end{document}